\documentstyle[11pt]{article}
\include{epsf}
\begin{document}
\thispagestyle{empty}
\begin{flushright}
MIT-CTP-2801 \\
cond-mat/9811139
\end{flushright}
\vskip5mm
\begin{center}
{ \LARGE A formula for the hopping expansion of  \\
8-vertex models coupled to an external field } 
\vskip13mm
Christof Gattringer
\vskip5mm
{\sl Massachusetts Institute of Technology \\
Center for Theoretical Physics \\
77 Massachusetts Avenue \\
Cambridge MA 02139, USA}
\vskip20mm
\begin{abstract}
We study a generalized 8-vertex model where the vertices are coupled to a 
locally varying field. We rewrite the partition function as an integral over
Grassmann variables. In this form it is possible to explicitly evaluate
all terms of the hopping expansion. Applications of the resulting formula, 
in particular
its relation to 2-D lattice field theories with fermions are discussed.
\end{abstract}
\vskip3mm
{\sl To appear in: International Journal of Modern Physics A}
\vskip10mm
\end{center}
\newpage
\section{Introduction} 
Rewriting physical systems in different representations is a powerful tool 
in theoretical physics. Alternative representations of a system can stress 
different aspects and allow new insights into its physical behavior. Sometimes
a model is more accessible to numerical methods in a rewritten form or can
even
be solved explicitly. 

Here we study a generalized version of the 
8-vertex model \cite{FaWu69,FaWu70,Ba82}, where now the vertices couple to an 
external field. This generalized model encompasses a large class of 
physically interesting systems such as spin models with locally varying 
coupling, polymers in an external field and in particular 2-D lattice
field theories with fermions.  
The model first will be represented as an integral over Grassmann
variables and in this representation one then can explicitly evaluate 
all terms of the hopping expansion. The resulting expression is a new 
representation for the partition function of the original model in terms
of loops. In this new form it is possible to make connection to 2-dimensional 
lattice field theories with Wilson fermions \cite{Ga98b}. The formula 
e.g.~can be used
to considerably simplify the hopping expansion of the lattice 
fermion determinant in an external scalar field, 
since it establishes that a large class of contributions (all terms with 
multiply occupied links) gets cancelled. Furthermore, our formula
allows to explicitly integrate out the external field and to obtain simple 
loop representations for e.g.~the Gross-Neveu model 
(see Section 5 for a brief discussion of these applications).

The article is organized as follows. In the next section we introduce the 
model and discuss its use for polymer systems and 2-D statistical models.  
Section 3 contains the reformulation of the partition function as an integral 
over Grassmann variables and the hopping expansion. In Section 4 the 
traces over the hopping generators are computed and the final expression for
the hopping expansion is presented. The article closes with 
a brief discussion 
of the properties and applications of our expansion formula. 
\vspace{5mm}

\section{Definition of the generalized 8-vertex model} 
\begin{figure}
\centerline{
\epsfysize=4.5cm \epsfbox[ 0 0 256 155 ]{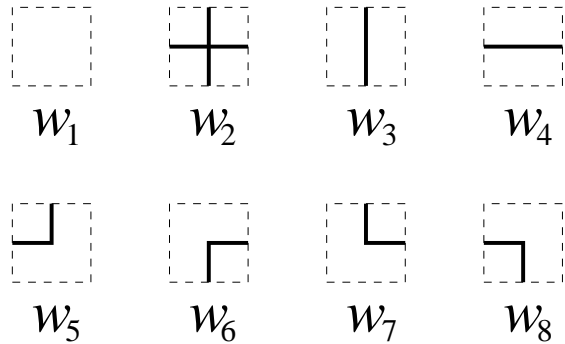}}
\protect\caption{ {The eight vertices (tiles) and their weights $w_i$.}
\label{tiles}}
\end{figure}

We analyze a generalized version of the 8-vertex model 
where now the vertices are coupled to a locally 
varying external field $B_\mu(x)$. The standard 8-vertex model 
\cite{FaWu69,FaWu70,Ba82} can be viewed 
as a model of 8 quadratic tiles (vertices) and 
each of them is assigned a weight $w_i\;  (i = 1, ... 8)$ 
(compare Fig.~\ref{tiles}).
Consider now a lattice $\Lambda$ which we assume to be a 
finite, rectangular piece of $\mbox{Z\hspace{-1.3mm}Z}^{2}$
(the generalization to e.g.~a torus is straightforward).
A {\it tiling} of this lattice
is a covering of $\Lambda$ with the tiles such that on each site of $\Lambda$
we place one of our tiles with the centers of the tiles sitting on the sites.
The set ${\cal T}$ of {\it admissible tilings} 
is given by those arrangements of 
tiles where the black lines on the tiles never have an open end (for sites
at the boundary this implies that not all 8 tiles can be used there). The 
partition function of the standard 8-vertex model is the sum over all 
admissible tilings $t \in {\cal T}$
and the Boltzmann weight for a particular tiling $t$  
is given by the product of the weights $w_i$ for all tiles used in 
the tiling $t$.

In our generalization of the model we now add an additional, local structure.
To each link $(x,\mu), \; \mu = 1,2$ of 
$\Lambda$ we assign a real- or complex-valued 
field $B_\mu(x), \; \mu = 1,2$. The {\it contour} $C(t)$ of a tiling $t$ 
is defined to be the set of all links of $\Lambda$ which are occupied by
black lines from the tiles. Since we allow only admissible tilings,
each link is either occupied or empty and `half-occupied' links do not occur.
The partition function of the generalized model is now given by
\begin{equation}
Z \; = \; \sum_{t \in {\cal T}} \; \; \prod_{i = 1}^8 {w_i}^{n_i(t)} \! \!
\prod_{(x,\mu) \in C(t)} B_\mu(x) \; .
\label{form1}
\end{equation}
By $n_i(t)$ we denote the abundance of tile Nr.~$i$ in a given tiling
$t$. Note that the contour $C(t)$ is simply the set of links occupied
by black lines and does not have an orientation. So in case 
the field $B_\mu$ is 
chosen complex, a link variable $B_\mu(x)$ is always counted as it is, and 
no complex conjugation is implied. 

A second (less general) form of the model is obtained by assigning a scalar
field $\varphi(x)$ to all sites of $\Lambda$ and setting 
\[
B_\mu(x) \; = \; \sqrt{\varphi(x)} \sqrt{\varphi(x+\hat{\mu})} \; ,
\]
where $\hat{\mu}$ denotes the unit vector in direction $\mu$. In case
one of the $\varphi$ is negative $B_\mu$ becomes complex, a case which we
explicitly included above.
The partition function now reads
\begin{equation}
Z \; = \; \sum_{t \in {\cal T}} \; \prod_{i = 1}^8 {w_i}^{n_i(t)} \!
\prod_{x \in S(t)} \varphi(x) \; .
\label{form2}
\end{equation}
Here $S(t)$ denotes the set of all sites occupied by the tiling $t$.
When a site $x$ is occupied by tile Nr.~2, this site is counted twice
giving a factor $\varphi(x)^2$. In case $x$ is occupied by tile Nr.~1,
$x \notin S(t)$ and the factor is 1. For all other tiles $x$ is 
counted once and the factor is $\varphi(x)$.
\\

The generalized 8-vertex model in its two forms (\ref{form1}), (\ref{form2})
encompasses several interesting physical systems
(for a detailed discussion of the corresponding models at trivial 
external field see \cite{Ba82} and references therein). For example
by setting the parameters in formulation 
(\ref{form1}) to $w_1 = w_2 = ... = w_8 = 1$ and the external fields to 
$B_\mu(x) = e^{-2 J_\mu(x) / kT}$, the model describes the Ising model 
with locally varying coupling $J_\mu(x)$ for links $(x,\mu)$. 
The set $C(t)$ has the interpretation of a Peierls contour which 
separates patches of up and down spins. Ising type models
with next to nearest neighbor terms can be obtained by chosing different 
values for the weights $w_i$. The model with locally varying couplings 
can furthermore be viewed as a generating functional for $2n$-point
functions in Ising type models: Differentiating $Z$ in the form 
(\ref{form1}) with
respect to the link variables on a connected path produces the 
2-point function for two spins sitting at the endpoints of the path
(for $2n$-point functions use a net of paths).

In its form (\ref{form2}), the generalized vertex model can be used to 
describe the physics of loop gases and polymers (compare
e.g.~\cite{Pr78,RyHe82,KaThHeRy83,KoPr85,BlSh90,Wi86}) 
in an external field $\varphi$.

Most remarkable, however, is the fact that our model is related to 
2-D lattice field theories with Wilson fermions. For staggered fermions
it has long been known, that a polymer representation can be found
\cite{RoWo84,KaSchTh85,Mo89,ArFoGa98}. Due to their spinor structure, 
a similar
map for Wilson fermions \cite{Wi74} is more complicated. Based on the
hopping expansion which will be discussed in the next section, it was shown
in \cite{Ga98b} that 2-D Wilson fermions in a scalar background 
field $\varphi(x)$ are equivalent to our model in its form (\ref{form2})
(see Section 5 for a brief discussion of the argument used in 
\cite{Ga98b}).

\section{Grassmann representation and hopping expansion}

The first step in computing the hopping expansion is finding a representation 
of the partition function through a Grassmann integral. For the standard 
8-vertex model this problem has been solved \cite{Sa80,FrSrSu80,It82}. The 
central idea is to assign to each site $x$ of the lattice 4 Grassmann 
variables $\eta_{+1}(x), \eta_{-1}(x), \eta_{+2}(x), \eta_{-2}(x)$ and 
integrate a Boltzmann factor with an appropriately chosen action $S[\eta]$ 
to obtain 
\begin{equation}
Z = (-1)^{|\Lambda|} \! \int \prod_{x \in \Lambda} 
d\eta_{-2}(x) d\eta_{+2}(x) \; d\eta_{-1}(x)
d\eta_{+1}(x) e^{S[\eta]} .
\label{pathint}
\end{equation}
Here $|\Lambda|$ denotes the size of the lattice. The
action is a quadratic form in the $\eta_i$  and contains two types of terms.
Hopping terms which are products of Grassmann variables at neighboring sites
create link elements, i.e.~the black lines on our tiles. In addition, 
quadratic terms based on only one site are needed to saturate the Grassmann
integral. These terms enable the different moves the lines perform
on the different tiles, e.g.~a corner on tiles 5-8, or keep going straight 
on tiles 3,4. For 
an explicit discussion of the Grassmann representation for the standard 
8-vertex model see \cite{Sa80}.

For the generalized model in the external field $B_\mu(x)$ only the hopping
terms have to be modified such that the links are furnished with their 
corresponding fields. The action $S = S_h + S_m + S_c$
producing the generalized 8-vertex model
then consists of hopping, monomer and corner terms 
\begin{eqnarray}
S_h & = & \sqrt{w_2} \sum_{x \in \Lambda} \; \Big[ 
B_1(x) \; \eta_{+1}(x) \eta_{-1}(x + \hat{1}) 
\; 
\nonumber \\
& & \hskip8mm + \; B_2(x) \; \eta_{+2}(x) \eta_{-2}(x + \hat{2}) \Big] \; , 
\nonumber \\
S_m & = & - \frac{1}{\sqrt{w_2}} \sum_{x \in \Lambda} \Big[ w_3  
\eta_{-1}(x) \eta_{+1}(x) + w_4  \eta_{-2}(x) \eta_{+2}(x) \Big] ,
\nonumber \\
S_c & = & - \frac{1}{\sqrt{w_2}} \sum_{x \in \Lambda} \Big[ w_5 \;
\eta_{+1}(x) 
\eta_{-2}(x)  +  
w_6 \; \eta_{+2}(x) \eta_{-1}(x)  
\nonumber \\
& & \hskip8mm
 + \; w_7 \; \eta_{-2}(x) \eta_{-1}(x) + 
w_8 \; \eta_{+2}(x) \eta_{+1}(x) \Big] \; . 
\label{grassact}
\end{eqnarray}
The boundary conditions are open, i.e.~hopping terms that would lead out of
our rectangular lattice are omitted. Inserting (\ref{grassact}) into 
the path integral, expanding the exponent termwise
and integrating the fields
along the lines of \cite{Sa80} establishes the representation (\ref{pathint})
for the partition function (\ref{form1}).

We remark, that the action (\ref{grassact}) is 
independent of $w_1$. However, it turns out, that for the Grassmann 
representation (\ref{pathint}), (\ref{grassact}) with a quadratic action
to work, the weights $w_i$ have to obey the free fermion 
condition \cite{FaWu69}
\begin{equation}
w_1 w_2 \; + \; w_3 w_4 \; = \; 
w_5 w_6 \; + \; w_7 w_8 \; .
\label{freecond}
\end{equation} 
Thus $w_1$ can be computed from the independent parameters $w_i, i > 1$.
Note that for the case of a {\it trivial external field} (all $B_\mu(x) = 1$),
the free fermion condition is a sufficient condition for finding an explicit
solution. This is most easily seen by first constructing the Grassmann 
representation (which always can be done when (\ref{freecond}) holds) and
then performing a Fourier transformation of the Grassmann variables.
The Fourier transformation diagonalizes the action and the free energy 
can be readily computed. However, when the fields $B_\mu(x)$ are 
chosen non-trivially, Fourier transformation produces a convolution of the 
external field and does no longer diagonalize the action. 
\\

Here we analyze the hopping expansion for the generalized 8-vertex model.
The next step is to anti-symmetrize the action by subtracting from
the above expression (\ref{grassact}) the same terms
but with reversed order of the Grassmann variables and dividing by 2.
It is convenient to denote the resulting action using matrix notation.
We order the Grassmann variables in a vector ($T$ denotes transposition)
\begin{equation}
\eta(x) \; = \; 
\Big( \eta_{+1}(x), \eta_{-1}(x), \eta_{+2}(x), \eta_{-2}(x) \Big)^T \; ,
\end{equation}
and define the $4\times 4$ matrices
$P_{\pm 1}$ and $P_{\pm 2}$
\begin{eqnarray}
P_{+1} (i,j) \equiv \sqrt{w_2} 
\delta_{i,1} \delta_{j,2} & \; , & 
P_{-1} (i,j)  \equiv  - \sqrt{w_2}  \delta_{i,2} \delta_{j,1} , 
\nonumber \\ 
P_{+2} (i,j) \equiv  \sqrt{w_2}  
\delta_{i,3} \delta_{j,4} & \; , \; & 
P_{-2} (i,j) \equiv - \sqrt{w_2} \delta_{i,4} \delta_{j,3} . 
\nonumber \\
\label{pmatrix}
\end{eqnarray}
They obey $P_\nu^T = - P_{-\nu}$. We also define
\[
\mu \; = \; \frac{1}{\sqrt{w_2}}
\left( \begin{array}{cccc} 
  0  & -w_3 & -w_8 & +w_5 \\ 
+w_3 &  0   & -w_6 & -w_7 \\ 
+w_8 & +w_6 &  0   & -w_4 \\ 
-w_5 & +w_7 & +w_4 &  0 
\end{array} \right) \; ,
\]
with
\[ 
\mu^{-1} \; = \; \frac{\sqrt{w_2}}{w_1 w_2}
\left( \begin{array}{cccc} 
 0   & -w_4 & +w_7 & -w_6 \\ 
+w_4 &  0   & +w_5 & +w_8 \\ 
-w_7 & -w_5 &  0   & -w_3 \\ 
+w_6 & -w_8 & +w_3 & 0 
\end{array} \right) \; .
\]
We remark that for writing $\mu^{-1}$ in this  
form the free fermion condition
(\ref{freecond}) was used. The determinant of $\mu$ is given by
$\det \mu = w_1^2$ and $\mu^T = -\mu$. 

Using the matrix notation the action reads
\[
S[\eta] \; = \; \frac{1}{2} \sum_{x,y} \eta(x)^T K(x,y) \eta(y) \; ,
\]
with the kernel $K$ consisting of 2 parts $K = -M + R$, where
\[
M(x,y) = \mu \; \delta_{x,y} \; \; , \; \; 
R(x,y) = \sum_{\nu = \pm 1}^{\pm 2} \; B_\nu(x) P_\nu \;
\delta_{x+\hat{\nu},y} \; ,
\]
and we have defined
\[
B_{-\nu}(x) \; = \; B_\nu(x-\hat{\nu}) \; , \; \; \; \nu \; = \; 1,2 \; .
\]
It has to be stressed, that here the link variable is {\it not} complex
conjugated when hopping in negative $\nu$-direction. This is different 
from lattice gauge theory (see e.g.~\cite{MoMu94}) where the link variables 
are conjugated for backward hopping. This difference indicates that for 
lattice field theories with fermions coupled to a vector field
a different structure emerges.

The partition function can now be written as a Pfaffian, and since by 
construction the kernel $K$ of the action is anti-symmetric, the Pfaffian 
is given by the root of the determinant of $K$. An overall factor 
$\det[-M] = w_1^2$ can be extracted and the rest is expanded in powers 
of the hopping matrix giving
\begin{eqnarray}
Z & = & (-1)^{|\Lambda|} \; \int D \eta \; e^{\frac{1}{2} \eta^T K \eta}
\; = \; (-1)^{|\Lambda|} \; \mbox{Pf} \; K \nonumber \\
& = & 
(-1)^{|\Lambda|} \; \sqrt{ \det[-M + R]} \nonumber \\
& = & (-1)^{|\Lambda|} \sqrt{ \det[-M]}\sqrt{ \det[1 - M^{-1} R]} 
\nonumber \\
& = &
(-1)^{|\Lambda|} w_1^{|\Lambda|} \; 
e^{- \frac{1}{2} \sum_{n=1}^\infty \frac{1}{n} \mbox{Tr} H^n},
\nonumber \\
\label{hopexp}
\end{eqnarray}
where we introduced the hopping matrix
\begin{equation}
H(x,y) \; = \; M^{-1} R(x,y) \; = \; \frac{1}{w_1} 
\sum_{\nu = \pm 1}^{\pm 2} \; B_\nu(x) \Gamma_\nu \;
\delta_{x+\hat{\nu},y} \; ,
\label{hoppingmatrix}
\end{equation}
with 
$
\Gamma_\nu \; = \; w_1 \mu^{-1} P_\nu$, 
or explicitly
\begin{eqnarray}
\Gamma_{+1} = \left( \begin{array}{cccc} 
 0 &  0   & 0  & 0 \\ 
 0 & +w_4 & 0  & 0 \\ 
 0 & -w_7 & 0  & 0 \\ 
 0 & +w_6 & 0  & 0 
\end{array} \right) & , & 
\Gamma_{-1} = \left( \begin{array}{cccc} 
 +w_4 & 0 & 0  & 0 \\ 
 0   & 0 & 0  & 0 \\ 
 +w_5 & 0 & 0  & 0 \\ 
 +w_8 & 0 & 0  & 0 
\end{array} \right) \; ,
\nonumber \\
\Gamma_{+2} = \left( \begin{array}{cccc} 
 0 & 0 & 0  & +w_7 \\ 
 0 & 0 & 0  & +w_5 \\ 
 0 & 0 & 0  & 0   \\ 
 0 & 0 & 0  & +w_3 
\end{array} \right) & , &
\Gamma_{-2} = \left( \begin{array}{cccc} 
 0 & 0 & +w_6   & 0 \\ 
 0 & 0 & -w_8  & 0 \\ 
 0 & 0 & +w_3   & 0 \\ 
 0 & 0 & 0     & 0 
\end{array} \right) \; .
\nonumber
\end{eqnarray}
The series in the exponent of (\ref{hopexp}) converges for $||H|| < 1$. 
By choosing suitable values for the weights $w_i$ and the fields $B_\mu(x)$
this can be achieved. Note however, that this restriction is only a technical 
assumption due to the expansion used here. The final result is simply a 
finite polynomial in the $w_i$ and $B_\mu(x)$ and the restriction of the
variables can be lifted then.

Due to the Kronecker deltas in (\ref{hoppingmatrix})
the traces Tr$H^n$ reduce to products of matrices $\Gamma_\nu$ supported 
on closed loops $l$. Since on the rectangular lattice all closed loops
have even length, traces over odd powers of $H$ vanish. 
For even powers we obtain
\begin{equation}
\mbox{Tr}\;  H^{2k} \; = \; \left( \frac{1}{w_1} \right)^{2k}
\sum_{x \in \Lambda} \sum_{l \in {\cal L}^{(2k)}_x} \; \prod_{(y,\nu) \in l} B_\nu(y) \;
\mbox{Tr} \prod_{\mu \in l} \Gamma_\mu \; .
\label{htrace}
\end{equation}
By ${\cal L}^{(2k)}_x$ we denote the set of all closed, connected 
loops of length
$2k$ and base point $x$. The first product picks up a factor $B_\nu(y)$
whenever the loop $l \in {\cal L}_x^{(2k)}$ runs through the 
link $(y,\nu)$. The last term is simply the trace of the product
of matrices $\Gamma_\mu$ as they appear along the loop. By explicit 
calculation one can show that 
\[
\Gamma_{\pm \mu} \; \Gamma_{\mp \mu} \; = \; 0 \; , \; \; \; \; \; \; \; 
\mu = 1,2 \; .
\]
This implies that whenever the loop $l$ turns around at a site and runs 
back on its last link the contribution of this {\it back-tracking} loop 
vanishes. Thus the set ${\cal L}^{(2k)}_x$ only contains closed, connected, 
non back-tracking loops of length $2k$ with base point $x$. For each loop
both possible orientations have to be taken into account. The loops in 
${\cal L}^{(2k)}_x$ can self-intersect and retrace parts of their contour
or even their whole contour.

\section{Computing the traces and final result}
\begin{figure}
\centerline{
\epsfysize=7.5cm \epsfbox[ 0 0 643 416 ] {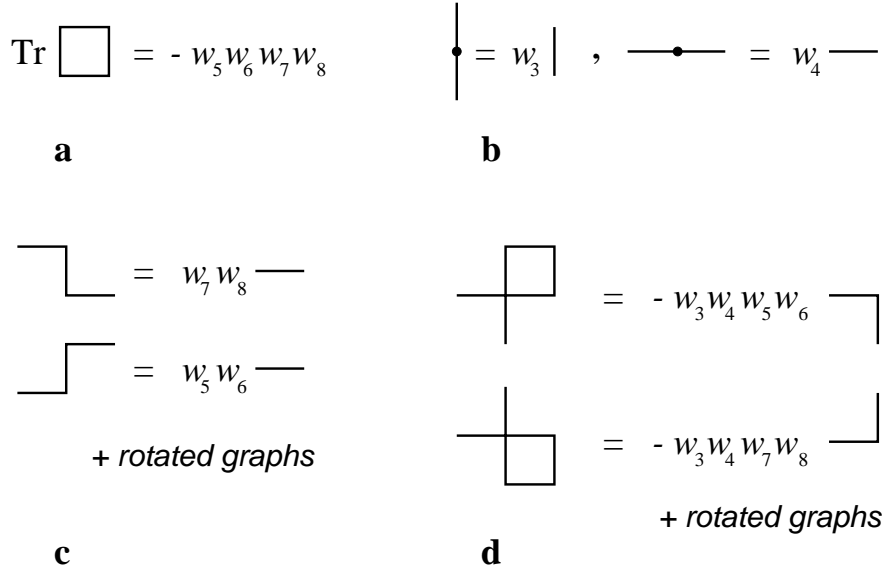}}
\vskip3mm
\protect\caption{ {The four elements used for the 
computation of the traces:
a: basic loop, b: telescope rule, c: kink rule, d: intersection rule.}
\label{elements}}
\end{figure}

\noindent
In this section we evaluate the traces over the matrices $\Gamma_\mu$. The
loops $l \in {\cal L}^{(2k)}_x$ are decomposed into 4 simple elements
(depicted in Fig.~\ref{elements}) and
algebraic relations for products of $\Gamma_\mu$ corresponding to these 
elements can be used to give a constructive procedure for determining 
the traces Tr$\prod_{\mu \in l} \Gamma_\mu$ (compare also \cite{GaJaSe98} 
where the case of the Ising model in this expansion was discussed in more 
detail).

The first of the four elements, {\it basic loop}, is the closed
loop around a single plaquette
(compare Fig.~\ref{elements}.a). One finds by direct evaluation
\begin{equation}
\mbox{Tr} \; \Gamma_{\mu} \Gamma_{\nu} 
\Gamma_{-\mu} \Gamma_{-\nu} \; =
\; - w_5 w_6 w_7 w_8 \; , 
\label{basic}
\end{equation}
where $\mu,\nu = \pm1, \pm2 \; , \;  
\mu \neq \pm \nu$.
Thus for the basic loop around a 
single plaquette we always obtain $-1$ times the product of the 
weights for the 4 tiles showing a corner (compare Fig.~\ref{tiles}),
independent of the starting point and the orientation.

The second element is referred to as {\it telescope rule}. 
The telescope rule can also be shown
by explicit evaluation and reads
\[
\Gamma_{\pm 1} \Gamma_{\pm 1} \; = \; w_4 \Gamma_{\pm1} \; \; \; \; , 
\; \; \; \; \Gamma_{\pm 2} \Gamma_{\pm 2} \; = \; w_3 \Gamma_{\pm2}
\; .
\]
The geometrical interpretation of these two identities is simple. A piece of
loop can be shrunk or stretched in horizontal (vertical) direction
and a factor $w_4$ ($w_3$) has to be collected (compare the graphical 
representation in Fig.~\ref{elements}.b).

The third rule is the {\it kink rule} depicted in Fig.~\ref{elements}.c.
Algebraically it reads
\begin{eqnarray}
\Gamma_{\pm \mu} \Gamma_{\pm \nu} \Gamma_{\pm \mu} & = &
w_5 w_6 \; \Gamma_{\pm \nu}
\; , \; \; \; \mu,\nu = 1,2 \; , \; \mu \neq \nu \; ,
\nonumber \\
\Gamma_{\pm \mu} \Gamma_{\mp \nu} \Gamma_{\pm \mu} & = &
w_7 w_8 \; \Gamma_{\pm \nu} 
\; , \; \; \; \mu,\nu = 1,2 \; , \; \mu \neq \nu \; .
\nonumber 
\end{eqnarray}
This rule, together with the telescope rule allows to remove superfluous 
kinks or corners in a loop (or sub-loop) and to reduce this loop (sub-loop)
to rectangular form which then can be shrunk to a loop (sub-loop)
around a single plaquette using the telescope rule. One simply replaces 
a piece of loop by a simpler piece (i.e.~one with less corners) which 
can be constructed using telescope and kink rule (see \cite{GaJaSe98} for
a detailed description of the procedure). During this process one has to 
collect all emerging factors $w_i$. In case the loop we started with 
contained no sub-loop, this procedure will reduce it to the basic loop
and using (\ref{basic}) the result for the trace can be read off. 
\\

In case the loop contains sub-loops, we first reduce a sub-loop which has 
only one self-intersection (such a loop always exists) to a sub-loop around
a single plaquette. The sub-loop can then be removed by the fourth rule,
the {\it intersection rule}. It states, that when removing a sub-loop, in 
addition to the corner and monomer factors, a factor 
of $-1$ has to be collected.
The algebraic expression reads (for the graphical representation see 
Fig.~\ref{elements}.d) 
\begin{eqnarray}
\Gamma_{\pm \mu} \Gamma_{\pm \mu} \Gamma_{\pm \nu} \Gamma_{\mp \mu} 
\Gamma_{\mp \nu} \Gamma_{\mp \nu} & = & - w_3w_4w_5w_6 \; \Gamma_{\pm \mu}
\Gamma_{\mp \nu} \; ,
\nonumber \\
\Gamma_{\pm \mu} \Gamma_{\pm \mu} \Gamma_{\mp \nu} \Gamma_{\mp \mu} 
\Gamma_{\pm \nu} \Gamma_{\pm \nu} & = & - w_3w_4w_7w_8 \; \Gamma_{\pm \mu}
\Gamma_{\pm \nu} \; ,
\nonumber
\end{eqnarray}
for $\mu,\nu = 1,2 \; , \; \mu \neq \nu$.
As outlined above, and discussed in more detail in \cite{GaJaSe98}, 
the four elements can be combined to compute the trace in a constructive way. 
After collecting all monomer, corner and intersection factors, we find the
result
\begin{equation}
\mbox{Tr} \; \prod_{\mu \in l} \; \Gamma_\mu 
\; = \; - (-1)^{s(l)} \prod_{i=3}^8 
w_i^{n_i(l)} \; .
\label{gtrace}
\end{equation}
Here $s(l)$ denotes the number of self-intersections the loop $l$ has. 
The exponents $n_i(l)$ give the numbers for the abundance of the line 
elements as they are depicted in Fig.~\ref{tiles}. E.g.~when the loop 
changes from heading east to heading north at a site, it picks up a 
factor of $w_5$, and similarly for the other tiles. 
Note that the loops $l$ in the hopping expansion appear as an ordered set
of instructions for the directions the loop takes as it hops from one site
to the next. The element corresponding to tile Nr.~2 with weight $w_2$
(compare Fig.~\ref{tiles}) does not appear. We also remark, that the 
result (\ref{gtrace}) is independent of the orientation of the loop.   

Inserting (\ref{gtrace}) and (\ref{htrace}) into (\ref{hopexp})
one finds
\begin{equation}
Z =   
(-1)^{|\Lambda|} w_1^{|\Lambda|} \; \exp \Bigg(
\frac{1}{2} \sum_{k=1}^\infty \frac{{w_1}^{-2k}}{2k} 
\sum_x \! \sum_{l \in {\cal L}^{(2k)}_x} (-1)^{s(l)} \!
\prod_{(y,\nu) \in l} B_\nu(y) \;
\prod_{i=3}^8 
w_i^{ni(l)} \Bigg). 
\label{baseresult}
\end{equation}
The final step is to eliminate the explicit summation over the base points. 
When doing so a minor subtlety has to be discussed. Consider first a 
loop which runs through its contour only once. Each of the $2k$ lattice points
it visits can serve as the base point producing a factor $2k$ which cancels
the corresponding factor in (\ref{baseresult}). If now a loop $l$ completely 
iterates its contour $I(l)$-times, then there are only $2k/I(l)$ different 
choices for a base point and a factor $1/I(l)$ remains. The final result is
\begin{equation}
Z \; = \; 
(-1)^{|\Lambda|} w_1^{|\Lambda|} \times  
\exp \left( \sum_{l \in {\cal L}} \frac{(-1)^{s(l)}}{I(l)} 
\Big( \frac{1}{w_1} \Big)^{|l|}
\prod_{(x,\nu) \in l} B_\nu(x) \;
\prod_{i=3}^8 
w_i^{n_i(l)} \right). 
\label{finalresult}
\end{equation}
Here the sum runs over the set ${\cal L}$ of all closed, 
non back-tracking loops of arbitrary length. Each loop is included with
only one of its two possible orientations leading to the cancellation of the 
factor 1/2 which appeared in the last expression. By $|l|$ we denote the length
of the loop $l$ and $I(l)$ is the number of iterations of its complete contour.
We remark, that (\ref{finalresult}) does not explicitly contain $w_2$, but
this weight is related to the other weights through the free fermion condition
(\ref{freecond}).

\section{Discussion of the result}
\noindent
Several aspects of the result (\ref{finalresult}) should be discussed.
When viewing vertex models as models of Peierls contours for Ising-type
spin models the exponent in (\ref{finalresult}) is proportional to the free 
energy in terms of loops. For the standard Ising model a representation of
this type is given in \cite{LaLi80}. Equation (\ref{finalresult}) generalizes 
this results to 8-vertex models coupled to an external field which also
includes the case of Z\hspace{-1.3mm}Z$_2$-spin models with locally varying 
couplings. 
\\

The expression (\ref{finalresult}) is also remarkable from an algebraic
point of view.
In principle, it is possible to expand the exponent in (\ref{finalresult})  
in a power series of ${w_1}^{-2k}$. The coefficients in this series
can be expressed in terms of products of loops $l \in {\cal L}$. In turn one 
would like to identify these loops with the contours $C(t)$ which occur
for the tilings $t$ in the original formulation of the model. The loops 
$l \in {\cal L}$, 
however, can occupy links several times and also taking products of
these loops will create multiply occupied links while the contours
$C(t)$ only have empty or singly occupied links. When analyzing some 
terms of this expansion, one finds however,
that due to the self-intersection factor
$(-1)^{s(l)}$ all contributions where links are multiply occupied get 
cancelled. Only terms without multiply occupied links survive and these
correspond to the contors $C(t)$. Our equation (\ref{finalresult})
establishes a remarkable feature of this expansion, namely
that the cancellation of terms with multiply occupied links is independent 
of the corner and monomer weights. When applying the formula to fermionic
lattice field theories (see below) this cancellation of all terms with 
multiply occupied links implies a considerable simplification of the hopping
expansion for the fermion determinant.
\\

Probably the most interesting application of Eq.~(\ref{finalresult}) 
is its use when relating fermionic lattice field theories to 
vertex models. An example of this type of application of our formula
Eq.~(\ref{finalresult}) is discussed in \cite{Ga98b}, where it is proven, 
that 2-D Wilson fermions in an external scalar field 
$\varphi(x)$ are equivalent  
to our model in its form (\ref{form2}) with $w_1 = 1, w_2 = 0,
w_3 = w_4 = \kappa, w_4 = ... \; w_8 = \kappa / \sqrt{2}$, where $\kappa$ 
is related to the mass parameter through $\kappa = (2+m)^{-1}$. For the
case of free Wilson fermions ($\varphi(x) = 1$) this was first shown by 
Scharnhorst \cite{Sch97} with different methods (for a presentation using 
the techniques of this paper see \cite{Ga98a}). \cite{Ga98b} extends this 
result to non-trivial background fields using a careful analysis of the 
hopping expansion for the Wilson fermions. Wilson fermions give rise to
a bilinear Grassmann action (instead of a quadratic form here) and the 
object to be expanded is the fermion determinant (instead of the Pfaffian).
The Dirac $\gamma$-matrices play the role of the hopping generators 
$\Gamma_\mu$ which we encounter here, and also for the Wilson fermions 
it is possible to compute all the traces which appear in the hopping 
expansion. The final step in the proof of equivalence is the identification
of the correct weights $w_i$ by comparing the hopping expansions for the
two models. The details of this mapping are given in \cite{Ga98b}.
Further results are obtained in \cite{Ga98b} by integrating out the scalar 
field, which can e.g.~be used to generate the Gross-Neveu model 
and map it to another vertex model (for similar mappings of the
Schwinger model and the 2-D Thirring model with Wilson fermions see 
\cite{Sch97,Sa91,Ga99}). 
Such mappings of lattice field theories onto vertex models 
(loop representation) are powerful tools, 
since often the numerical simulation of the vertex model is simpler than 
analyzing the original model and allows for higher precision (see 
e.g.~\cite{GaLaSa92} where this was demonstrated for the
case of the strongly coupled Schwinger model). In particular the loop 
expansion
of the fermion determinant developed here and in \cite{Ga98b} 
overcomes the fermion sign problem.
This permutation sign plagues fermionic systems in the Hamiltonian approach
\cite{ChWi99} as well as in the Grassmann path integral 
formulation \cite{Cr98} and
in a numerical treatment requires an exponentially increasing number of 
Monte Carlo configurations as the volume or inverse temperature are 
increased \cite{ChWi99}.
Currently we are working on an implementation of new numerical algorithms 
which are based directly on the loop representation \cite{GaGaSt99}.
\\
\\
{\sl Acknowledgement:} The author thanks Christian Lang, Klaus Scharnhorst 
and 
Uwe-Jens Wiese for discussions and remarks on the literature.

\end{document}